\documentclass{article}
\usepackage[utf8]{inputenc}
\usepackage[tbtags]{amsmath}
\usepackage{amsthm}
\usepackage{amssymb}
\usepackage[letterpaper, portrait, left=1in, right=1in]{geometry}
\usepackage{xcolor}
\usepackage{array}
\usepackage{comment}
\usepackage{enumerate}
\usepackage{graphicx}
\usepackage{authblk}
\usepackage{ulem}

\newcounter{comments}
\definecolor{Red}{rgb}{0.8,0,0}
\definecolor{Green}{rgb}{0.2,0.6,0.2}
\definecolor{Blue}{rgb}{0,0,0.8}

\newcommand{\new}[2][]{{#2}}

\newcommand{\delete}[1]{}

\allowdisplaybreaks
\DeclareMathOperator*{\expectation}{\mathbb{E}}
\DeclareMathOperator*{\prob}{\mathbb{P}}

\DeclareMathOperator{\vecx}{\mathbf{x}}

\newtheorem{theorem}{Theorem}[section]

\newtheorem{corollary}[theorem]{Corollary}
\newtheorem{conjecture}[theorem]{Conjecture}
\theoremstyle{definition}
\newtheorem{definition}[theorem]{Definition}
\newtheorem{remark}[theorem]{Remark}
\newtheorem{example}[theorem]{Example}

\title{Collectively canalizing Boolean functions} 
\author[1]{Claus Kadelka}
\author[2]{Benjamin Keilty}
\author[3]{Reinhard Laubenbacher}
\affil[1]{Department of Mathematics, Iowa State University, Ames, Iowa 50011}
\affil[2]{Department of Mathematics, University of Connecticut, Storrs, CT, 06269}
\affil[3]{Laboratory for Systems Medicine, Department of Medicine, University of Florida, Gainesville, FL 32610}

\begin{document}
\maketitle

\begin{abstract}
This paper studies the mathematical properties of collectively canalizing Boolean functions, a class of functions that has arisen from applications in systems biology. Boolean networks are an increasingly popular modeling framework for regulatory networks, and the class of functions studied here captures a key feature of biological network dynamics, namely that a subset of one or more variables, under certain conditions, can dominate the value of a Boolean function, to the exclusion of all others. These functions have rich mathematical properties to be explored. The paper shows how the number and type of such sets influence a function's behavior and define a new measure for the canalizing strength of any Boolean function. We further connect the concept of collective canalization with the well-studied concept of the average sensitivity of a Boolean function. The relationship between Boolean functions and the dynamics of the networks they form is important in a wide range of applications beyond biology, such as computer science, and has been studied with statistical and simulation-based methods. But the rich relationship between structure and dynamics remains largely unexplored, and this paper is intended as a contribution to its mathematical foundation. 
\end{abstract}

\section{Introduction}

One of the great advances of biology in the twentieth century is the discovery of genes and their regulatory relationships, now increasingly described as gene regulatory networks that are amenable to a description by dynamic mathematical models. Traditionally, this has been done using systems of ordinary differential equations, one per gene in the network, based on the view of the network as a biochemical reaction network, subject to constraints, such as preservation of mass. An alternative view, arguably closer to biological thinking, is to represent gene regulatory networks as similar to logical switching networks, popular in engineering, employing ON/OFF state representations instead of continuously varying concentrations, introduced by S. Kaufman\cite{kauffman1969metabolic,kauffman1974large}. This explains the increasing popularity of such models in biology, in particular, since in many cases detailed kinetic measurements are not readily available. A natural next question then is what the biological constraints are on the Boolean functions that occur in this way. 

In the 1940s, C. Waddington introduced the concept of canalization in developmental biology as an explanation for the surprising stability of developmental processes in the face of varying environmental conditions~\cite{waddington1942canalization}. S. Kauffman later adapted this concept and introduced a canalization concept for Boolean functions~\cite{kauffman2003random,kauffman2004genetic}. A multitude of studies have shown that Boolean networks composed of canalizing functions exhibit more ordered dynamics than random ones, resulting in, e.g., fewer and shorter attractors as well as lower sensitivities to perturbations~\cite{kauffman2003random,kauffman2004genetic,shmulevich2004activities,karlsson2007order,peixoto2010phase,kadelka2017influence,paul2020dynamics}. 

A canalizing function possesses at least one input variable such that, if this variable takes on a certain ``canalizing" value, then the output value of the function is already determined, regardless of the values of the remaining input variables. If this variable takes on another, non-canalizing, value, and there is a second variable with this same property, the function is $2$-canalizing. If $k$ variables follow this pattern, the function is $k$-canalizing~\cite{he2016stratification}, and the number of variables that follow this pattern is called the canalizing depth of the function~\cite{layne2012nested}. If the canalizing depth equals the number of inputs (i.e. if all variables follow the described pattern), the function is also called nested canalizing. 

It is straightforward to see that any Boolean function can be represented as a polynomial over the field with two elements, first exploited in \cite{laubenbacher2004computational}, to use tools from computational algebra for the inference of Boolean network models from experimental data. He and Macauley showed that any Boolean function can be written in a unique canonical form, from which the number of Boolean functions with a certain canalizing depth can be easily derived~\cite{he2016stratification}. In addition, explicit formulas for the number of various types of Boolean and multi-state canalizing and nested canalizing functions have also been found~\cite{just2004number,murrugarra2012number,li2013boolean,kadelka2017multistate}. Given the stringency of the definition of canalization it is not surprising that a random Boolean function in several variables is only rarely canalizing, let alone nested canalizing. It is thus remarkable that most functions found in published gene regulatory network models are indeed canalizing or even nested canalizing~\cite{harris2002model,daniels2018criticality,kadelka2020meta}, suggesting that the canalization property does indeed capture an important property of the logic of gene regulatory networks. 

Biological observations suggest that, in many cases, a given gene is regulated by a collection of other genes that jointly determine that gene's dynamics. Based on this phenomenon, less stringent definitions of canalization have been considered~\cite{bassler2004evolution,reichhardt2007canalization}. Most Boolean functions exhibit some degree of canalization in the sense that a few variables taking on certain ``canalizing" input values frequently suffice to determine the output of a function, regardless of the values of the remaining input variables. This phenomenon has been first described and studied by Bassler et al., and has been termed collective canalization~\cite{bassler2004evolution}. The amount of canalization a particular Boolean function exhibits is described by the set of numbers $P_k, k=0,1,\ldots,n-1$, which are the fraction of $k$-dimensional input sets that are collectively canalizing. Another way to think of these numbers is as the probability that the output of the Boolean function is already determined if $k$ randomly chosen inputs are fixed. Reichhardt and Bassler used results from group theory and isomer chemistry to classify all Boolean functions in $n$ variables based on the set of numbers $P_k, k=0,1,\ldots,n-1$ and parity symmetry (which describes if a Boolean function is symmetric, antisymmetric or not symmetric about its midpoint, i.e. if all inputs are flipped), and derived the number of different classes and the size of each class~\cite{reichhardt2007canalization}.

In this paper, we expand on this early work on collectively canalizing Boolean functions. In Section~\ref{sec:definitions}, we review some concepts frequently used in the analysis of Boolean functions and provide a mathematically rigorous definition of collective canalization. In Section~\ref{sec:canalizing_strength}, we investigate the set of numbers $P_k, k=0,1,\ldots,n-1$, described above, and combine these numbers into a single quantity that describes the canalizing strength of any Boolean function. In Section~\ref{sec:sensitivity}, we provide bounds for the average sensitivity of a Boolean function in terms of the set of numbers $P_k, k=0,1,\ldots,n-1$ connecting the concepts of average sensitivity and canalization. We conclude with a brief discussion in Section~\ref{sec:discussion}.

\section{Collective canalization}\label{sec:definitions}
In this section, we review some concepts and definitions, introduce the concept of \emph{canalization}, and  generalize it to \emph{collective canalization}, following earlier work by Reichhardt and Bassler~\cite{reichhardt2007canalization}. Throughout the paper, let $\oplus$ denote addition modulo $2$ when used in a polynomial, and the ``exclusive or" (XOR) function when used in a Boolean logical expression.

\begin{definition}
A Boolean function $f(x_1,\ldots,x_n)$ is essential in the variable $x_i$ if there exists an $\vecx \in \{0,1\}^n$ such that 
$$f(\vecx)\neq f(\vecx\oplus\, e_i),$$
where $e_i$ is the $i$th unit vector.
\end{definition}

\begin{definition}\label{canal}
A Boolean function $f(x_1,\ldots,x_n)$ is canalizing if there exists a variable $x_i$, a Boolean function $g(x_1,\ldots,x_{i-1},x_{i+1},\ldots,x_n)$ and $a,\,b\in\{0,\,1\}$ such that
$$f(x_1,x_2,...,x_n)= \begin{cases}
b,& \ \text{if}\ x_i=a\\
g(x_1,x_2,...,x_{i-1},x_{i+1},...,x_n),& \ \text{if}\ x_i\neq a
\end{cases}$$
In that case, we \new{call $x_i$ a \emph{canalizing variable} and }say that $x_i$ \emph{canalizes} $f$. \new[$x_i$]{More specifically, if $x_i$ receives the \emph{canalizing input value} $a$ it} {canalizes} $f$ to \new{the \emph{canalized output value} }$b$.
\par Some authors further require the function $g$ to be non-constant; in this paper, we do not impose this requirement. This is because when defining collectively canalizing functions, it makes sense to include constant $g$, and it is convenient to have our definition of canalizing functions here correspond to the definition of $1$-set canalizing functions in Definition~\ref{def:coll_canal}.
\end{definition}

\begin{definition}\label{def2.3}\cite{he2016stratification} 
A Boolean function $f(x_1,\ldots,x_n)$ is $k$-canalizing, where $1 \leq k \leq n$, with respect to the permutation $\sigma \in \mathcal{S}_n$, inputs $a_1,\ldots,a_k$ and outputs $b_1,\ldots,b_k$, if
\begin{equation*}f(x_{1},\ldots,x_{n})=
\left\{\begin{array}[c]{ll}
b_{1} & x_{\sigma(1)} = a_1,\\
b_{2} & x_{\sigma(1)} \neq a_1, x_{\sigma(2)} = a_2,\\
b_{3} & x_{\sigma(1)} \neq a_1, x_{\sigma(2)} \neq a_2, x_{\sigma(3)} = a_3,\\
\vdots  & \vdots\\
b_{k} & x_{\sigma(1)} \neq a_1,\ldots,x_{\sigma(k-1)}\neq a_{k-1}, x_{\sigma(k)} = a_k,\\
g\not\equiv b_k & x_{\sigma(1)} \neq a_1,\ldots,x_{\sigma(k-1)}\neq a_{k-1}, x_{\sigma(k)} \neq a_k,
\end{array}\right.\end{equation*}
where $g = g(x_{\sigma(k+1)},\ldots,x_{\sigma(n)})$ is a Boolean function on $n-k$ variables. When $g$ is not canalizing, the integer $k$ is the canalizing depth of $f$ (as in \cite{layne2012nested}). An $n$-canalizing function is also called nested canalizing function (NCF), and we define all Boolean functions to be $0$-canalizing.
\end{definition}

He and Macauley provided the following powerful stratification theorem. 

\begin{theorem}\cite{he2016stratification}\label{thm:he}
Every Boolean function $f(x_1,\ldots,x_n)\not\equiv 0$ can be uniquely written as 
$$f(x_1,\ldots,x_n) = M_1(M_2(\cdots (M_{r-1}(M_rp_C \oplus 1) \oplus 1)\cdots)\oplus 1)\oplus b,$$
where each $M_i = \prod_{j=1}^{k_i} (x_{i_j} \oplus a_{i_j})$ is a non-constant extended monomial, $p_C$ is the \emph{core polynomial} of $f$, and $k = \sum_{i=1}^r k_i$ is the \emph{canalizing depth}. Each $x_i$ appears in exactly one of $\{M_1,\ldots,M_r,p_C\}$, and the only restrictions are the following ``exceptional cases":
\begin{enumerate}
    \item If $p_C\equiv 1$ and $r\neq 1$, then $k_r\geq 2$;
    \item If $p_C\equiv 1$ and $r = 1$ and $k_1=1$, then $b=0$.
\end{enumerate}
When $f$ is a non-canalizing functions (i.e., when $k=0$), we simply have $p_C = f$.
\end{theorem}

Theorem~\ref{thm:he} shows that any Boolean function has a unique standard monomial form, in which the variables are partitioned into different layers based on their dominance. Any canalizing variable is in the first layer. Any variable that ``becomes" canalizing when excluding all variables from the first layer is in the second layer, etc. All remaining variables that never become canalizing are part of the core polynomial. The number of variables that ``become" eventually canalizing is the canalizing depth, and NCFs are exactly those functions where all variables ``become" eventually canalizing.

\begin{definition}\cite{kadelka2017influence}\label{def:layer}
The \emph{layer structure} of a Boolean function $f(x_1,\ldots,x_n)$ with canalizing depth $k$ is defined as the vector $(k_1,\ldots,k_r)$, where $r$ is the \emph{number of layers} and $k_i$ is the \emph{size of the $i$th layer}, $i=1,\ldots,r$. The layer structure follows directly from the unique standard monomial form of $f$ (Theorem~\ref{thm:he}).
\end{definition}

\begin{example}
The function $f(x_1,x_2,x_3)=x_1\land (x_2 \lor x_3)$ is nested canalizing with layer structure $(k_1=1,k_2=2)$. The unique standard monomial form is $f = M_1(M_2 \oplus 1)$, where $M_1 = x_1$ and $M_2 = (x_2\oplus1)(x_3\oplus1)$.
\end{example}

\begin{remark}\label{rem:layers}
\new{Variables in the same layer of a $k$-canalizing Boolean function may have different canalizing input values, however they canalize the function to the same output (i.e., have the same canalized output value). On the other hand, the outputs of two consecutive layers are distinct. Therefore, the number of layers of a $k$-canalizing function expressed as in Definition~\ref{def2.3} is simply one plus the number
of changes in the vector of canalized outputs, $(b_1,b_2,\ldots,b_k)$. More specifically, all variables in odd layers canalize $f$ to $b_1$, while all variables in even layers canalize $f$ to $b_1\oplus 1$.}
\end{remark}

\begin{definition}\label{def:coll_canal}
A Boolean function $f(x_1,\ldots,x_n)$ is $k$-set canalizing, where $0\leq k\leq n$, if there exists a permutation $\sigma \in \mathcal{S}_n$, inputs $a_1,\ldots, a_k \in\{0,1\}$ and an output $b\in\{0,1\}$ such that
$$f(x_1,x_2,...,x_n)=\begin{cases}
b,&(x_{\sigma(1)},\,x_{\sigma(2)},\ldots,x_{\sigma(k)})=(a_1,a_2,\ldots,a_k),\\
g(x_1,\ldots, x_n),&\text{otherwise.}
\end{cases}$$ 
In that case, the input set $C_k=\{(x_{\sigma(1)},a_1),(x_{\sigma(2)},a_2),...,(x_{\sigma(k)},a_k)\}$ \emph{(collectively) canalizes} $f$ (to $b$). 
\end{definition}

\begin{definition}
For $0\leq k\leq n$, the \emph{$k$-set canalizing proportion} of a Boolean function $f(x_1,\ldots,x_n)$, denoted $P_k(f)$, is defined as the proportion of $k$-sets $C_k$ from Definition~\ref{def:coll_canal}, which collectively canalize $f$.
\end{definition}

\begin{remark}\label{remark_canalizing}
These definitions imply the following.
\begin{enumerate}[(a)]
    \item A function $f$ is $k$-set canalizing if and only if $P_k(f)> 0$.
    \item For any function $f(x_1,\ldots,x_n)$, $P_n(f)\equiv 1$.
    \item Constant functions are the only $0$-set canalizing functions. That is, $P_0(f)=0$ except when $f$ is a constant function, in which case $P_0(f)=1$.
    \item Canalizing functions, as defined in Definition~\ref{canal}, are exactly the $1$-set canalizing functions.
    \item Consider the $n$-dimensional Boolean cube $B_n$, with vertices labeled according to $f$. $P_k(f)$ is the probability that any $(n-k)$-face of $B_n$ is constant.
\end{enumerate}
\end{remark}
\begin{example}
The function $f(x_1,x_2,x_3,x_4)=(x_1\lor x_2)\land (x_3\lor x_4)$ is not canalizing, $P_1(f) = 0$. However, $f$ is $2$-set canalizing because if $x_1=0$ and $x_2=0$, then $f\equiv 0$, regardless of the values of $x_3$ and $x_4$. Thus, $\{(x_1,0),(x_2,0)\}$ collectively canalizes $F$ to $0$. Similarly, $\{(x_3,0),(x_4,0)\}$ canalizes $f$ to $0$, while $\{(x_1,1),(x_3,1)\}, \{(x_1,1),(x_4,1)\}, \{(x_2,1),(x_3,1)\},$ and $\{(x_2,1),(x_4,1)\}$ canalize $f$ to $1$. Thus, $P_2(f) = \frac 6{24} = \frac 14$.
\end{example}

\section{Quantifying the canalizing strength of any Boolean function}\label{sec:canalizing_strength}
In this section we investigate the $k$-set canalizing proportion of various types of functions and use it to define the canalizing strength of any Boolean function, a measure which we argue more accurately resembles the biological concept of canalization. We begin by showing that the $k$-set canalizing proportion can never decrease in $k$. 

\begin{theorem}\label{thm_proportion_increasing}
Let $f(x_1,\ldots,x_n)$ be a Boolean function. Then for $1\leq k<n$,
$$P_{k-1}(f)\leq P_k(f)\leq \frac 12 (1+P_{k-1}(f)).$$
\end{theorem}

\textbf{Proof.} Let $[n] = \{1,2,\ldots,n\}$. Let $f(x_1,\ldots,x_n)$ be a Boolean function, and let $\mathcal{C}_k$ be the set of all $k$-input sets that collectively canalize $f$. For an input set $C=\{(x_{\sigma(1)},a_1),(x_{\sigma(2)},a_2),...,(x_{\sigma(k-1)},a_{k-1})\}$ with $|C| = k-1$, define an extended input set $C^*(\sigma(k),a_k) = \{(x_{\sigma(1)},a_1),(x_{\sigma(2)},a_2),...,(x_{\sigma(k-1)},a_{k-1}),(x_{\sigma(k)},a_k)\}$, where $\sigma(i)\neq \sigma(j)$ whenever $i\neq j$. Further, let $P_C(f)$ be the proportion of all possible extended input sets $C^*$ which collectively canalize $f$. Clearly, 
$$P_k(f)=\mathbb{E}[P_C(f)],$$
where the expectation is taken uniformly over all possible input sets $C$ with $|C|=k-1$.

\noindent Case 1, $C \in \mathcal{C}_{k-1}$: If $C$ already collectively canalizes $f$, then $P_C(f) = 1$.

\noindent Case 2, $C \not\in \mathcal{C}_{k-1}$: We will consider all $2(n-(k-1))$ extended input sets $C^*$ and show that $P_C(f) \leq \frac 12$.

\noindent Case 2a, $\exists j \in [n] - \{\sigma(1),\sigma(2),\ldots,\sigma(k-1)\}$ such that $C^*(j,0)$ and $C^*(j,1)$ both collectively canalize $f$ to the same output: This implies that $C$ already collectively canalizes $f$ but this contradicts $C \not\in \mathcal{C}_{k-1}$.

\noindent Case 2b, $\exists j \in [n] - \{\sigma(1),\sigma(2),\ldots,\sigma(k-1)\}$ such that $C^*(j,0)$ and $C^*(j,1)$ both collectively canalize $f$ to different output values: This implies that $C^*(j,0)$ and $C^*(j,1)$ are the only two choices for $C^*$ that collectively canalize $f$. Since there are $n-(k-1)$ choices for $j$ and each has two corresponding $C^*$, we have $P_C(f)=\frac 2{2(n-(k-1))}=\frac 1{n-k+1}\leq\frac 12$.

\noindent Case 2c, $\nexists j \in [n] - \{\sigma(1),\sigma(2),\ldots,\sigma(k-1)\}$ such that $C^*(j,0)$ and $C^*(j,1)$ both collectively canalize $f$: This implies that at most one of the two corresponding $C^*$ collectively canalizes $f$, thus $P_C(f) \leq \frac 12$.

\noindent By definition, $\mathbb{P}(C\in \mathcal{C}_{k-1}) = P_{k-1}(f)$. Therefore, conditioning on $C\in \mathcal{C}_{k-1}$ yields
\begin{align*}P_k(f) &= \mathbb{E}[P_C(f)]\\
&=\mathbb{P}(C\in \mathcal{C}_{k-1}) \mathbb{E}\big[P_C(f) \ \big| \ C\in \mathcal{C}_{k-1}\big]  + \mathbb{P}(C\not\in\mathcal{C}_{k-1})\mathbb{E}\big[P_C(f) \ \big| \ C\not\in \mathcal{C}_{k-1}\big]\\
&= P_{k-1}(f)\cdot 1+(1-P_{k-1}(f))\mathbb{E}\big[P_C(f) \ \big| \ C\not\in \mathcal{C}_{k-1}\big]
\end{align*}
Thus,
\begin{align*}
P_{k-1}(F)\leq P_k(F)&\leq P_{k-1}(F)+\big(1-P_{k-1}(F)\big)\frac 12 = \frac 12 \big(1+P_{k-1}(F)\big)
\end{align*}
\qed

\begin{corollary}\label{corollary_constant_P}
For a constant Boolean function $f(x_1,\ldots,x_n)$, $P_0(f)=P_1(f)=\ldots=P_n(f)=1$. If $f$ is not constant, then $P_k(f)\leq 1-\frac 1{2^k}$ for all $0\leq k<n$.
\end{corollary}
\textbf{Proof.} If $f(x_1,\ldots,x_n)$ is constant, then $P_0(f)=1$ and $P_n(f)=1$ by Remark~\ref{remark_canalizing}. Thus, by Theorem~\ref{thm_proportion_increasing}, $P_0(f) = P_1(f) = \ldots = P_n(f) = 1$.
\par If $f(x_1,\ldots,x_n)$ is not constant, then $P_0(f)=0=1-\frac 1{2^0}$. Proceed by induction and assume that $P_{k-1}(f)\leq 1-\frac 1{2^{k-1}}$ for some $k<n-1$. Then by Theorem~\ref{thm_proportion_increasing},
\begin{align*}
        P_k(f)&\leq \frac12(1+P_{k-1}(f)) \leq \frac12\left(1+1-\frac1{2^{k-1}}\right)= 1-\frac1{2^k}
\end{align*}\qed

\par In fact, we can show that the equality $P_k(f)=1-\frac1{2^k}$ only holds if $f$ is a special type of canalizing function, an NCF with exactly one layer (see Theorem~\ref{thm:he} and Definition~\ref{def:layer}). 
\new{However, we first prove a more general theorem that provides a formula for $P_k(f)$ for any NCF with known layer structure.}

\new{
\begin{theorem}\label{thm_any_ncfs}
If $f(x_1,\ldots,x_n)$ is a Boolean NCF with known layer structure $k_1,\ldots,k_r$, where $r$ is the number of layers and $k_1+\cdots +k_r = n$, then for all $0\leq k<n$, 
\begin{align*}
P_k(f)&= \frac{1}{\binom nk 2^k} \Bigg[ \binom{k_{r-1} + k_{r-3} + \ldots}{k - k_r - k_{r-2} - \ldots}\ + \\
&+ \sum_{i=1}^r \sum_{c\geq 1}  (2^{c}-1) \binom{k_i}{c} \sum_{d\geq 0} \binom{k_{i-2}+k_{i-4}+\ldots}{d} \binom{k_{i+1} + k_{i+2} + \cdots + k_r}{k - c - d - k_{i-1} - k_{i-3} - \ldots} 2^{k - c - d - k_{i-1} - k_{i-3} - \ldots} \Bigg].
\end{align*}
\end{theorem}

\textbf{Proof.} For any $k, 1\leq k\leq n$, let $\mathcal{C}_k$ be the set of all $2^k \binom nk$ input sets $$C_k=\{(x_{\sigma(1)},a_{\sigma(1)}),(x_{\sigma(2)},a_{\sigma(2)}),...,(x_{\sigma(k)},a_{\sigma(k)})\}$$ (remember: $\sigma(i) \neq \sigma(j)$ for $i\neq j$), each of which may collectively canalize a Boolean function $f(x_1,\ldots,x_n)$. To compute $P_k(f)$ for any nested canalizing function $f(x_1,\ldots,x_n)$ with known layer structure $k_1,\ldots,k_r$, where $r$ is the number of layers and $k_1+\cdots +k_r = n$, we will stratify all these sets as follows.

Let $f$ be written as in Theorem~\ref{thm:he}. In an abuse of notation, we express the extended monomials as $M_i = \{(x_{i_1},a_{i_1}),\ldots,(x_{i_{k_i}},a_{i_{k_i}})\}$, $i=1,\ldots,r$ to align with the notation for $C_k$. Further, we define $\bar M_i = \{(x_{i_1},a_{i_1}\oplus 1),\ldots,(x_{i_{k_i}},a_{i_{k_i}}\oplus 1)\}$ to be the ``negated" input sets where all variables in $M_i$ receive their non-canalizing input. We will now enumerate all the input sets of size $k$ that canalize $f$. For $k=1$, we have $C_k= \{(x_{\sigma(1)},a_{\sigma(1)})\}$ canalizes $f$ if and only if $(x_{\sigma(1)},a_{\sigma(1)}) \in M_1$. $M_1$ contains $k_1$ elements. Thus, $P_1(f) = \frac{k_1}{2n}$.


For any $k, 1\leq k\leq n$, we can stratify $\mathcal C_k$ into the following $r+2$ disjoint sets. For $i=1,\ldots,r$,
\begin{align*}
S_{i} =  \Big\{C_k \in \mathcal C_k\  \Big|& C_k \cap M_{i} \neq \emptyset \ \text{and}\\
& C_k \cap \bar M_{i+1-2j} = \bar M_{i+1-2j}\ \forall\ 1\leq j \leq \Big\lfloor \frac i2\Big\rfloor \ \text{and}\\
& C_k \cap M_{i-2j} = \emptyset\ \forall\ 1\leq j \leq \Big\lfloor \frac {i-1}2\Big\rfloor \Big\},\\
S_{r+1} = \Big\{C_k \in \mathcal C_k\  \Big|& C_k \cap \bar M_{r-2j} = \bar M_{r-2j}\ \forall\ 0\leq j \leq \Big\lfloor \frac {r-1}2\Big\rfloor\  \text{and}\\
& C_k \cap M_{r+1-2j} = \emptyset\ \forall\ 1\leq j \leq \Big\lfloor \frac {r}2\Big\rfloor \Big\},\\
S_{r+2} = \Big\{C_k \in \mathcal C_k\  \Big|& C_k \not\in S_i\ \forall 1\leq i \leq r+1\Big\}
\end{align*}

The first condition, $C_k \cap M_i \neq \emptyset$, in the definition of $S_i, i=1,\ldots,r$ ensures that $f$ is canalized in layer $i$, i.e., we know that at least one of the variables in layer $t$ receives its canalizing input. The second and the third condition ensure that layer $i$ is even reached in the nested evaluation of $f$. Layer $i$ can only determine the output of $f$ if all variables in those more important (i.e., outside) layers, which would lead to a different output, receive their non-canalizing input; due to Remark~\ref{rem:layers} these are exactly the variables in layers $i-1, i-3$, etc. Similarly, layer $i$ can only determine the output of $f$ if none of the variables in those more important (i.e., outside) layers, which lead to the same output (e.g. layers $i-2$, $i-4$, etc.), receive their canalizing input. Thus, the third condition ensures that the sets $S_i, S_{i-2}, S_{i-4}$, etc. are disjoint. Without this condition, some input sets in $S_i$ for $i\geq 3$ may also occur in e.g. $S_{i-2}$ because a variable in a more important layer may have already received its canalizing input.

If none of the variables of an NCF receive their canalizing input value, the NCF evaluates to $b_{n}\oplus 1$ (Definition~\ref{def2.3}). The set $S_{r+1}$ contains all those input sets $C_k$, for which this is the case. We require two conditions similar to the second and third condition in the previous paragraph. First, all variables in layers $r, r-2, r-4$, etc. must receive their non-canalizing input so that $f$ is not canalized to $b_n$ (Remark~\ref{rem:layers}). Second, to ensure the sets $S_{r+1}, S_{r-1}, S_{r-3}$, etc. are mutually disjoint, none of the variables in layers $r-1, r-3$, etc. are allowed to receive their canalizing input; they may however receive their non-canalizing input.

Altogether, an input set $C_k$ collectively canalizes $f$ if and only $C_k \in S_i$ for some $i=1,\ldots,r+1$. To find the $k$-set canalizing proportion, we sum up the magnitudes of the sets $S_i, i=1,\ldots,r+1$. This computation relies heavily on the multivariate hypergeometric distribution: there are $\prod_{i=1}^r \binom{k_i}{c_i}$ ways to draw (without replacement) $k$ out of $n$ variables such that $c_i$ variables are part of $M_i$ (i.e., layer $i$), for $i=1,\ldots,r$. 

Let $C_k \in S_i$ for some $1\leq i \leq r$. The first condition in the definition of $S_i$ implies $c_i \geq 1$. The second condition implies that we need $c_j = k_j$ for $j = i-1, i-3$, etc. and that each variable in layers $i-1, i-3$, etc. must receive its non-canalizing input. Otherwise, $C_k \not\in S_i$. That means there is only one choice for the input value in $C_k$ associated with each variable in these layers. The third condition implies that while $c_j$ can be freely chosen for $j=i-2, i-4$, etc., there is again only one choice for the input value associated with any variable in these layers that is also part of $C_k$, namely the non-canalizing input. On the contrary, for all variables in layers $i+1, i+2, \ldots, r$ that are part of $C_k$, there are two choices for the associated input value, $0$ and $1$. Finally, for those $c_i\geq 1$ variables that are part of $C_k$ and of layer $i$, there is a total of $2^{c_i} - 1$ combined possible choices for the input values associated with these variables, as only at least one variable must receive its canalizing input. Thus, we have for $1\leq i\leq r$,

\begin{align*}
|S_i| &= \sum \Big[\prod_{j=1}^r \binom{k_j}{c_j}\Big] \Big[\prod_{m=i+1}^r 2^{c_m}\Big] (2^{c_i}-1),
\end{align*}
where the sum is taken over all $(c_1,\ldots,c_r)$ such that 
\begin{enumerate}
\item $c_1+c_2+\cdots+c_r = k$,
\item $0\leq c_j\leq k_j$ for all $1\leq j\leq r$,
\item $c_i \geq 1$ (first condition in the definition of $S_i$), 
\item $c_j = k_j$ for all $j=i-1,i-3$, etc. (second condition in the definition of $S_i$).
\end{enumerate}
With Vandermonde's identity, $\binom{a+b}{c} = \sum_{i=1}^c \binom{a}{i}\binom{b}{c-i}$, this simplifies to 
\begin{align*}
|S_i| &= \sum_{c\geq 1}  (2^{c}-1) \binom{k_i}{c} \sum_{d\geq 0} \binom{k_{i-2}+k_{i-4}+\ldots}{d} \binom{k_{i+1} + k_{i+2} + \cdots + k_r}{k - c - d - k_{i-1} - k_{i-3} - \ldots} 2^{k - c - d - k_{i-1} - k_{i-3} - \ldots}.
\end{align*}
By definition, for $n\geq 0$ we have $\binom{n}{m} = 0$ if $m>n$ and if $m<0$. Thus, the previous formula is correct without specific upper bounds for $c$ and $d$. For implementation purposes, the correct bounds would be 
\begin{align*}
c &= \max\big(1, k - n + k_i\big),\ldots, \min\big(k_i, k - k_{i-1} - k_{i-3} - \ldots\big),\\
d &= \max\big(0,k - n - c + k_i + k_{i-2} + k_{i-4} + ...\big),\ldots, \min\big(k_{i-2} + k_{i-4} + ..., k - c - k_{i-1} - k_{i-3} - \ldots\big).
\end{align*}
Similarly, we obtain
\begin{align*}
|S_{r+1}| &= \sum\prod_{j=1}^r \binom{k_j}{c_j} = \binom{k_{r-1} + k_{r-3} + \ldots}{k - k_r - k_{r-2} - \ldots}
\end{align*}
where the sum  is taken over all $(c_1,\ldots,c_r)$ such that 
\begin{enumerate}
\item $c_1+c_2+\cdots+c_r = k$,
\item $0\leq c_j\leq k_j$ for all $1\leq j\leq r$,
\item $c_j = k_j$ for all $j=r,r-2$, etc. (first condition in the definition of $S_{r+1}$).
\end{enumerate}

\qed

}

\begin{theorem}\label{thm_ncfs_with_one_layer}
If $f(x_1,\ldots,x_n)$ is a Boolean NCF with exactly one layer, then $P_k(f)=1-\frac1{2^k}$ for all $0\leq k<n$. Further, if for some $f(x_1,\ldots,x_n)$ with $n\geq 3$, there exists a $k$, $0<k<n$ such that $P_k(f)=1-\frac1{2^k}$, then $f$ is a NCF with exactly one layer.
\end{theorem}

\textbf{Proof.} 
Let $f$ be a Boolean NCF with exactly one layer. \new{$P_k(f)=1-\frac1{2^k}$ directly follows from Theorem~\ref{thm_any_ncfs} with $k_1 = n$. However, we provide an alternative, direct proof here. }By \new[Thm. 4.5 in \cite{he2016stratification}]{Theorem~\ref{thm:he}}, there exist $\alpha_1, \ldots, \alpha_n, \beta \in \{0,1\}$ such that $f$ can be uniquely written in standard monomial form,
$$f(x_1,\ldots,x_n) = \beta + \prod_{i=1}^n (x_{i} + \alpha_{i}).$$
Let $C_k=\{(x_{\sigma(1)},a_{\sigma(1)}),(x_{\sigma(2)},a_{\sigma(2)}),...,(x_{\sigma(k)},a_{\sigma(k)})\}$ be a randomly chosen input set of size $k, 0<k<n$, as in Definition~\ref{def:coll_canal}. Then, $C_k$ collectively canalizes $f$ to $\beta$ if $\prod_{i=1}^k (x_{\sigma(i)} + \alpha_{i}) = 0$, i.e. if $\alpha_{\sigma(i)} \neq a_{\sigma(i)}$ for some $1\leq i\leq k$. We have $P(\alpha_{\sigma(i)} = a_{\sigma(i)}) = \frac 12$ for all $1\leq i\leq k$ and thus, due to independence, 
$$P_k(f) = \prob\left(\exists i \in \{1,\ldots,k\} : \alpha_{\sigma(i)} \neq a_{\sigma(i)}\right) = 1 - \prob\left(\forall i \in \{1,\ldots,k\} : \alpha_{\sigma(i)} = a_{\sigma(i)}\right) = 1-\frac1{2^k}.$$ 
Further by Remark~\ref{remark_canalizing}, $P_0(f) = 0 = 1-\frac1{2^0}$ as $f$ is not constant.

To prove the second part, let $f$ be a Boolean function on $n\geq3$ variables such that $P_k(f)=1-\frac1{2^k}$ for some $0<k<n$. By Theorem~\ref{thm_proportion_increasing}, $P_k(f)\leq \frac12(1+P_{k-1}(f))$. Thus,
\begin{align*}
        1-\frac1{2^k}&\leq\frac12(1+P_{k-1}(f))\\
        1-\frac1{2^{k-1}}&\leq P_{k-1}(f)
\end{align*}
However, by Corollary~\ref{corollary_constant_P}, $P_{k-1}(f)\leq 1-\frac1{2^{k-1}}$, so in fact $P_{k-1}(f)=1-\frac1{2^{k-1}}$. Iteratively, we get $P_1(f)=\frac12$. 

Consider any variable $x_i$. If both $x_i=0$ and $x_i=1$ canalize $f$ to the same value, then $f$ is a constant function and $P_1(f)=1$, a contradiction. On the other hand, If both $x_i=0$ and $x_i=1$ canalize $f$ to different values then no other variable can canalize $f$, thus $P_1(f)=\frac1n$, contradicting $P_1(f)=\frac12$ for $n\geq 3$. Thus, only $x_i = 0$ or $x_i = 1$ can canalize $f$ for $n\geq 3$ and thus $P_1(f) \leq \frac 12$.

In order for $P_1(f)=\frac12$, we need every variable $x_i$ to canalize $f$ to the same value $b \in \{0,1\}$ for exactly one input $a_i$. Thus, we can express $f$ in standard monomial form \new[(Thm. 4.5 in \cite{he2016stratification})]{(Theorem~\ref{thm:he})},
$$f = b + \prod_{i=1}^n (x_{i} + \bar a_{i}),$$
and deduce that $f$ is an NCF with exactly one layer. \qed
\vspace{2ex}

Theorem~\ref{thm_ncfs_with_one_layer} provides maximal values for the $k$-set canalizing proportion of non-constant functions. The following example provides minimal values.

\begin{example}\label{ex_xor_minimum}
For $b \in \{0,1\}$, let $f(x_1,\ldots,x_n) = x_1 \oplus x_2 \oplus \ldots \oplus x_n \oplus b$ be the parity function. Then, $P_k(f) = 0$ for all $0\leq k < n$ because in any case knowledge of all inputs is required to determine the output.
\end{example}

Given maximal and minimal values for the $k$-set canalizing proportion of non-constant functions, we are now in a position to define a new robustness measure for any Boolean function with $n\geq 2$ inputs as a combination of the $k$-set canalizing proportions for all $k$ with $0<k<n$.  

\begin{definition}\label{can_strength}
The \textit{canalizing strength} of a non-constant Boolean function $f(x_1,\ldots,x_n)$ with $n\geq 2$ is defined as 
$$c(f) = \frac{1}{n-1}\sum_{k=1}^{n-1} \frac{2^k}{2^k-1}P_k(f) \in [0,1].$$
\end{definition}

\begin{example}
For $b \in \{0,1\}$, let $f(x_1,\ldots,x_n) = x_1 \oplus x_2 \oplus \ldots \oplus x_n \oplus b$ be the parity function as in Example~\ref{ex_xor_minimum}. Then, $c(f) = 0$, highlighting that the output of the parity function can only be determined when the values of all inputs are known.

On the other hand, if $f$ is a nested canalizing function with exactly one layer, f.e. the AND function $f(x_1,\ldots,x_n) = x_1x_2\cdots x_n$, then with Theorem~\ref{thm_ncfs_with_one_layer}, \begin{align*}c(f) &= \frac{1}{n-1}\sum_{k=1}^{n-1} \frac{2^k}{2^k-1}(1-\frac1{2^k})\\
&= \frac{1}{n-1}\sum_{k=1}^{n-1} 1\\
&= 1.
\end{align*}
\end{example}

\begin{remark}\label{remark_canalizing_strength}
The weights in Definition \ref{can_strength} are chosen such that (i) $c(f) \in [0,1]$ for any non-constant Boolean function and (ii) $c(f)=1$ for the ``most" canalizing functions (NCFs with exactly one layer), irrespective of $n$. This results however in $c(f) > 1$ for constant functions $f$. 

Alternatively, one could define the canalizing strength of a Boolean function \new{using} the unweighted average,
\new{$$c_{\text{unweighted}}(f) = \frac{1}{n-1}\sum_{k=1}^{n-1} P_k(f).$$}
This definition would ensure that any function (even constant ones) possesses a canalizing strength in $[0,1]$ \new{and that the canalizing strength can be directly interpreted as the probability that a randomly chosen subset of inputs determines the output, no matter the values of all other inputs, where the number of randomly chosen inputs is drawn uniformly from $1$ to $n-1$.} On the other hand, only Definition \ref{can_strength} ensures that the canalizing strength can be \new{readily} compared between functions with varying numbers of inputs (as the canalizing strength of any maximally canalizing function\new{, irrespective of the number of inputs,} is fixed at $1$), \new{and can thus} serve as a measure for the closeness to ``perfect" canalization (NCF with 1 layer \new{with $c(f) = 1$}). \new{We believe that the weighted definition of the canalizing strength is therefore more useful.} 
\end{remark}

\new{
\begin{remark}\label{remark_canalizing_strength_complexity}
The time required to calculate the exact value of all $k$-set canalizing proportions, $P_k(f)$, (and thus the canalizing strength) increases exponentially in the number of variables. Boolean functions may exhibit some level of symmetry.  In this context, two variables are part of the same symmetry group if their values can be interchanged without affecting the value of the function. Knowledge of the symmetry groups can reduce the time required to calculate the $k$-set canalizing proportions as only one member of each group needs to be considered. However, the problem of identifying the symmetry groups of a Boolean function is known to be NP-hard~\cite{stearns2019symmetry}. For this reason, we hypothesize there exists no algorithm that computes the canalizing strength of any Boolean function in polynomial time. For practical applications, Monte Carlo methods can yield approximate values of the $k$-set canalizing proportions and the canalizing strength for functions with any number of variables.
\end{remark}}

The next example highlights how the canalizing strength coincides more closely with the biological concept of canalization than, for instance, the canalizing depth or a simple binary measure of the presence/absence of canalizing variables.

\begin{example}\label{ex:canalizing_strength_explicit_example}
The function $f(x_1,\ldots,x_5) = x_1 \wedge \big(\oplus_{i=2}^5 x_i\big)$ is canalizing in $x_1$ (canalizing depth = 1). We have $P_1(f) = 0.1, P_2(f) = 0.2, P_3(f) = 0.3, P_4(f) = 0.5$, resulting in $c(f) \approx 0.336$. On the other hand, \new{the threshold function} $g(x_1,\ldots,x_5) = \big((x_1+\cdots+x_5)>1\big)$ is not canalizing  (canalizing depth = 0), $P_1(g) = 0$. However, $P_2(g) = 0.25, P_3(g) = 0.5, P_4(g) = 0.75$, resulting in $c(g) \approx 0.426$, which is larger than the canalizing strength of the canalizing function $f$.
\end{example}

This example highlights that functions that are canalizing in the traditional sense (Definition~\ref{canal}) may have a lower canalizing strength than functions not canalizing in the traditional sense. We thus investigated the distribution of the $k$-set canalizing proportion and the canalizing strength for different types of functions.

\begin{definition}(see f.e. \cite{shmulevich2004activities})
A random Boolean function $f(x_1,\ldots,x_n)$ with \emph{bias} $p$ can be generated by flipping a $p$-biased coin $2^n$ times and accordingly filling in the truth table. The bias is \textit{not} a property of an individual Boolean function; rather, it is a property of the underlying probability space. Setting $p=\frac12$ yields a uniform distribution on all Boolean functions $f: \{0,\,1\}^n\to\{0,\,1\}$.
\end{definition}

\begin{theorem}\label{thm_expectation_P_all_biased}
For a given bias $p\in(0,1)$ and for $0\leq k \leq n$, we have $$\expectation[P_{k}(f)]=  \left(1-p\right)^{2^{n-k}} + p^{2^{n-k}}.$$

In particular, when the expectation is taken uniformly over all $f\colon\{0,\,1\}^n\to\{0,\,1\}$ (i.e., in the unbiased case of $p = \frac 12$), $$\expectation[P_{k}(f)]=\frac1{2^{2^{n-k}-1}}.$$
\end{theorem}


\textbf{Proof.} Let $f$ be a Boolean function in $n$ variables, sampled uniformly at random from the space of $p$-biased Boolean functions. By Remark~\ref{remark_canalizing}e, the $k$-set canalizing proportion, $P_k(f)$, is the probability that an $(n-k)$-face of the $n$-dimensional Boolean cube, with vertices labeled according to $f$, is constant. Each $(n-k)$-face has $2^{(n-k)}$ vertices and there are two possible constants, $0$ and $1$, which are taken on with probability $1-p$ and $p$, respectively. Thus,
$$\expectation[P_{k}(f)] = \left(1-p\right)^{2^{n-k}} + p^{2^{n-k}}.$$ \qed

\begin{figure}
    \centering
\includegraphics[width=0.49\textwidth]{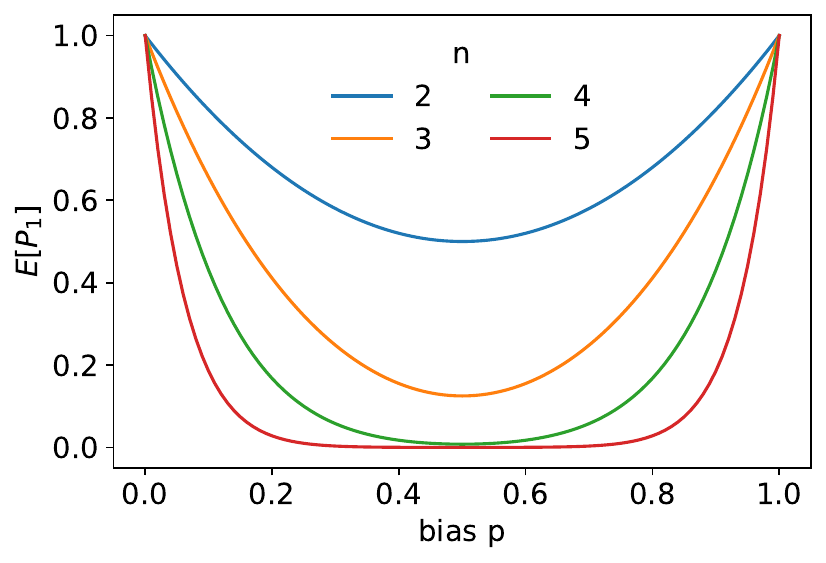} 
\includegraphics[width=0.49\textwidth]{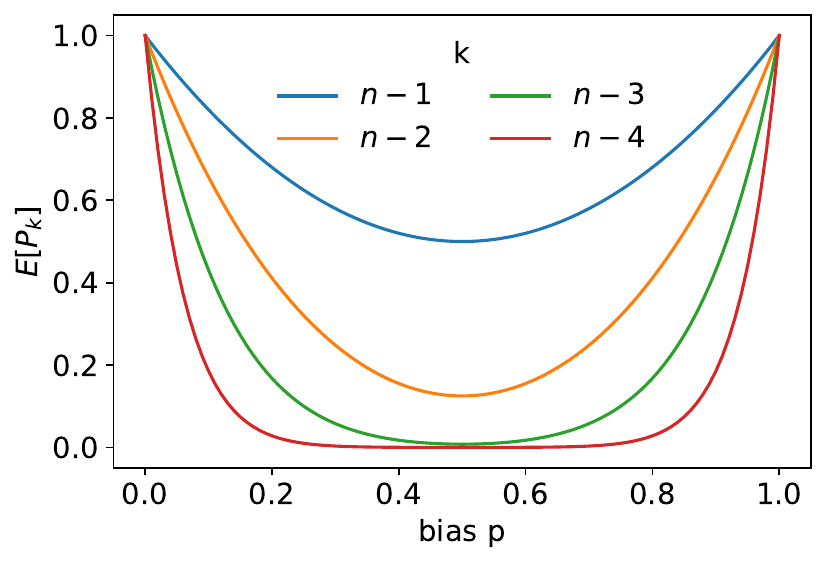} 

\caption{(A) Probability that a random input canalizes a function ($\mathbb{E}[P_1]$) for different sampling biases and different numbers of inputs ($n$). (B) Expected $k$-set canalizing proportion ($\mathbb{E}[P_k]$)  for different sampling biases and different values of $k \in \{n-4,n-3,n-2,n-1\}$.} 
    \label{fig:bias_vs_EP}
\end{figure}

Figure~\ref{fig:bias_vs_EP} highlights the implications of Theorem~\ref{thm_expectation_P_all_biased}. Unbiased functions ($p=0.5$) exhibit the lowest $k$-set canalizing proportion and thus the lowest canalizing strength, irrespective of $k$ or the number of inputs, $n$. Increased absolute bias leads to increased canalization.

\begin{figure}
    \centering
\includegraphics[width=0.49\textwidth]{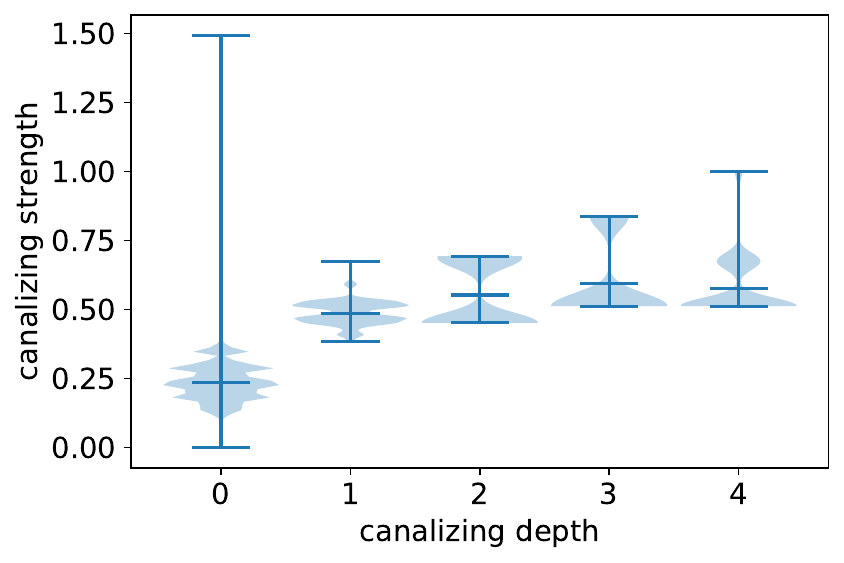} 
\includegraphics[width=0.49\textwidth]{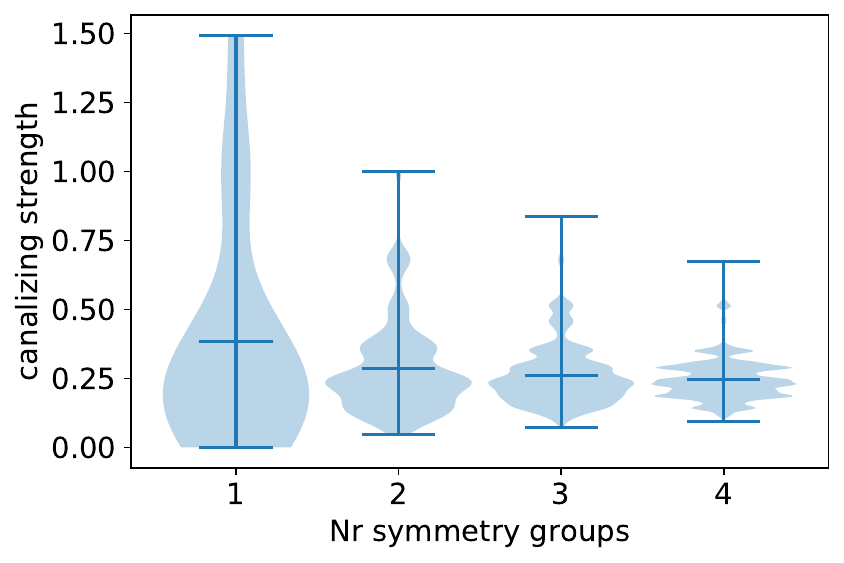} 

\caption{Distribution of the canalizing strength for all $2^{2^4} = 65,536$ Boolean functions in $n=4$ variables with a fixed (A) the canalizing depth, (B) number of symmetry groups. Horizontal dark lines depict the respective maximal, mean and minimal value (top to bottom).}
    \label{fig:canalizing_strength}
\end{figure}

An analysis of all Boolean functions in $n=4$ variables revealed that, on average, functions with more canalizing variables have a higher canalizing strength (Figure~\ref{fig:canalizing_strength}A). There are, however, strong variations and this result holds only in the average, as highlighted by Example~\ref{ex:canalizing_strength_explicit_example}. Functions in $n=4$ variables with canalizing depth $3=n-1$ all contain some non-essential variables, which explains their increased canalizing strength. In addition, functions with higher symmetry levels (that is, fewer symmetry groups) possess, on average, higher canalizing strengths, again with strong variations (Figure~\ref{fig:canalizing_strength}B). Note that the ``most" canalizing functions, NCFs with one layer, are all completely symmetric (that is, they have one symmetry group). 

Theorem~\ref{thm_expectation_P_all_biased} also directly yields the following corollary. 

\begin{corollary}\label{corollary_canalizing_strength_limit}
For any bias $p\in(0,1)$, the expected canalizing strength of randomly chosen Boolean functions approaches $0$ as the number of variables increases,
$$\mathbb{E}[c(f)] \underset{n \rightarrow \infty}{\longrightarrow} 0.$$
\end{corollary}
\textbf{Proof.} 
By Definition~\ref{can_strength}, Theorem~\ref{thm_expectation_P_all_biased} and linearity of the expectation, we have
\begin{align*}
    \mathbb{E}[c(f)] &= 
\frac 1{n-1} \sum_{k=1}^{n-1} \frac{2^k}{2^k-1} \left(\left(1-p\right)^{2^{n-k}} + p^{2^{n-k}}\right)\\
&\leq \frac 1{n-1} \sum_{k=1}^{n-1} 4\max\left(1-p,p\right)^{2^{n-k}}\\
&\leq \frac 4{n-1} \sum_{k=1}^{n-1} \max\left(1-p,p\right)^{n-k}\\
&\leq \frac 4{n-1} \frac{1}{1-\max\left(1-p,p\right)} \underset{n \rightarrow \infty}{\longrightarrow} 0.
\end{align*}\qed

An interesting, related question is the following: We know that the set of all \textit{canalizing} functions is very small compared to all Boolean functions. That is, $$\mathbb{P}\big(P_1(f) > 0 \ \big| \ f\colon\{0,\,1\}^n\to\{0,\,1\}\big) \underset{n \rightarrow \infty}{\longrightarrow} 0.$$
Similarly, we know that all Boolean functions except for the parity function and its conjugate have some constant edge in their hypercube representation. That is, all but two Boolean functions are $(n-1)$-set canalizing and 
$$\mathbb{P}\big(P_{n-1}(f) > 0  \ \big| \ f\colon\{0,\,1\}^n\to\{0,\,1\}\big) \underset{n \rightarrow \infty}{\longrightarrow} 1.$$
But what happens ``in between"? More precisely: For which $k$, does $$\mathbb{P}\big(P_{k}(f) > 0  \ \big| \ f\colon\{0,\,1\}^n\to\{0,\,1\}\big) \underset{n \rightarrow \infty}{\longrightarrow} 1?$$
The following, quite intuitive corollary provides some answers.


\begin{corollary}\label{thm_probability_proportion_limit}
For any bias $p\in(0,1)$ and any integer $k>0$,
$$\lim_{n\to\infty}\mathbb{P}\left(P_{n-k}(f)>0 \ \big|\ f\colon\{0,\,1\}^n\to\{0,\,1\}\right)\neq 0$$
while
$$\lim_{n\to\infty}\mathbb{P}\left(P_{k}(f)>0 \ \big|\ f\colon\{0,\,1\}^n\to\{0,\,1\}\right)=0.$$
\end{corollary}

\textbf{Proof.} By Theorem~\ref{thm_expectation_P_all_biased},
$$\expectation[P_{n-k}(f)]=  \left(1-p\right)^{2^{k}} + p^{2^{k}} > 0,$$
irrespective of $n$. This directly yields the first part of the corollary.
\par To prove the second part, we realize that all possible values of $P_k(f)$ are by definition fractions. There are $\binom nk 2^{k}$ different input sets, which contain $k$ out of $n$ variables. Thus, $P_k(f)>0$ implies $P_k(f)\geq \frac 1{\binom nk 2^k}$.
\par Now, assume $$\lim_{n\to\infty}\mathbb{P}\left(P_{k}(f)>0 \ \big|\ f\colon\{0,\,1\}^n\to\{0,\,1\}\right)=r\neq 0.$$
This implies 
$$\expectation[P_k(f)]\geq r\cdot \frac1{\binom nk 2^k}.$$ 
We can express $\binom nk$ as a polynomial in $n$ with degree $k$ and leading coefficient $\frac 1{k!}$, and get
$$\lim_{n\to\infty} n^k\expectation[P_k(f)]\geq \lim_{n\to\infty} r\cdot \frac{n^k}{\binom nk 2^k}= r\cdot\frac{k!}{2^k} > 0.$$
However, by Theorem~\ref{thm_expectation_P_all_biased} we have for any $p\in (0,1)$ that
$$\lim_{n\to\infty}n^k\expectation[ P_k(f)]=\lim_{n\to\infty} n^k\left( \left(1-p\right)^{2^{n-k}} + p^{2^{n-k}} \right)=0$$
by l'Hôpital's rule. This is a contradiction, which completes the proof.\qed

\section{Canalization and average sensitivity}\label{sec:sensitivity}
The average sensitivity, introduced in \cite{cook1986upper}, measures how sensitive the output of a function is to input changes, and constitutes one of the most studied properties of a Boolean function~\cite{li2013boolean,shmulevich2004activities,boppana1997average}. Thus far, average sensitivity and canalization were two distinct concepts. In this section, we derive bounds for the average sensitivity of a Boolean function in terms of the $k$-set canalizing proportions, allowing us to connect these two concepts.

\begin{definition}
The \emph{sensitivity} of a Boolean function $f(x_1,\ldots,x_n)$ at a vector $\mathbf x \in \{0,1\}^n$ is defined as the number of Hamming neighbors of $\mathbf x$ with a different function value than $f(\mathbf x)$. That is,
$$S(f,\mathbf x) = \sum_{i=1}^n \chi[f(\mathbf x) \neq f(\mathbf x \oplus e_i)].$$
\end{definition}

\begin{definition}
The \emph{average sensitivity} of a Boolean function $f(x_1,\ldots,x_n)$ is the expected value of $S(f,\mathbf x)$. Assuming a uniform distribution of $\mathbf x$,
$$S(f) = \mathbb{E}[S(f,\mathbf x)] = \frac 1{2^n} \sum_{\mathbf x \in \{0,1\}^n} \sum_{i=1}^n \chi[f(\mathbf x) \neq f(\mathbf x \oplus e_i)].$$
\end{definition}

\begin{definition}
Assuming a uniform distribution of $\mathbf x$, the \emph{normalized average sensitivity} of a Boolean function $f(x_1,\ldots,x_n)$ is
$$s(f)=\frac 1n S(f) = \frac 1n \sum_{i=1}^n \expectation\big[f(\mathbf{x})\oplus f(\mathbf{x}\oplus e_i)\big] = \frac 1{n2^n} \sum_{\mathbf x \in \{0,1\}^n} \sum_{i=1}^n \chi[f(\mathbf x) \neq f(\mathbf x \oplus e_i)].$$
\end{definition}

\begin{theorem}\label{thm_sensitivity_bounds}
For any Boolean function $f\colon\{0,\,1\}^n\to\{0,\,1\}$, and for any integer $0< k\leq n$, $$\frac 1{2^{k-1}}(1-P_{n-k}(f))\leq s(f)\leq 1-P_{n-k}(f).$$
\end{theorem}

\textbf{Proof.} 
We prove the left inequality using a geometric argument. By Remark~\ref{remark_canalizing}, $P_k(f)$ is the probability that an $(n-k)$-face of the $n$-dimensional Boolean cube $B_n$ is constant, where the vertices of $B_n$ are labeled according to $f$. Thus, $1-P_{n-k}(f)$ is the probability that a $k$-face is \textit{not} constant. Similarly, $s(f)$ is exactly the probability that a $1$-face (i.e., an edge) of $B_n$ is not constant. Let $H$ be a $k$-face of $B_n$ where $f$ is not constant. Any vertex in $H$ has $k$ edges that are part of $H$ and $H$ possesses $k2^{k-1}$ total edges. Since $f$ is not constant on $H$, there is at least one vertex in $H$ where $f$ takes on a different value. Thus, $H$ possesses at least $k$ non-constant edges, and by summing over all (constant and non-constant) $k$-faces we get
\begin{align*}
    s(f) &\geq (1-P_{n-k}(f))\cdot \frac{k}{k2^{k-1}} + P_{n-k}(f)\cdot 0\\
    &= \frac{1}{2^{k-1}}(1-P_{n-k}(f))
\end{align*}

The right inequality is a direct consequence of Theorem~\ref{thm_proportion_increasing}.

\qed

Theorem~\ref{thm_sensitivity_bounds} provides bounds for the average sensitivity of a Boolean function given only some of its canalizing proportions, or in terms of Graph Theory, given only the proportion of monochromatic higher-dimensional sides of a Boolean cube, we provide upper and lower bounds for the number of monochromatic ($1$-dimensional) edges. Further, the $k=1$ case in Theorem~\ref{thm_sensitivity_bounds} directly yields the following trivial result, relating the normalized average sensitivity and the $(n-1)$-set canalizing proportion. 

\begin{corollary}\label{theorem_sensitivity_P_equality}
$s(f)=1-P_{n-1}(f)$ for any Boolean function $f\colon\{0,\,1\}^n\to\{0,\,1\}$.
\end{corollary}

The trivial upper bound provided by Theorem~\ref{thm_sensitivity_bounds} is a very weak result. We believe the establishment of a good, general upper bound is an elusive problem. One substantially better upper bound, for which we have no proof, is given by the following conjecture.

\begin{conjecture}\label{conj}
For any Boolean function $f\colon\{0,\,1\}^n\to\{0,\,1\}$, and for any integer $0< k\leq n$, $$s(f)\leq 1-\sqrt[k]{P_{n-k}(f)}.$$
\end{conjecture}

\new{
\begin{remark}
To get an intuition how tight the bounds provided by Theorem~\ref{thm_sensitivity_bounds} and Conjecture~\ref{conj} are, we classified all $65,536$ Boolean functions in $n=4$ variables based on their normalized average sensitivity, i.e., based on $1-P_{n-1}(f)$. For each class of functions with fixed normalized average sensitivity, we then determined the minimal and the maximal value of $P_{n-k}(f)$ for $k=3$~(Table~\ref{table_tight_bounds}A) and for $k=2$~(Table~\ref{table_tight_bounds}B). We used these values to derive the best lower and the best upper bound that Theorem~\ref{thm_sensitivity_bounds} and Conjecture~\ref{conj} provide for functions with a specific normalized average sensitivity. 

When only $P_{n-3}(f) = P_{1}(f)$ is known (that is, only the number of canalizing variables; Table~\ref{table_tight_bounds}A), the lower bound is as good as it can be except for one class, functions with an average sensitivity of $0.1875$. These are however degenerated functions with only three essential variables, which are all canalizing, e.g., $f(x_1,x_2,x_3,x_4) = x_1 \wedge x_2 \wedge x_3$. When $P_{n-2}(f) = P_{2}(f)$ is known (Table~\ref{table_tight_bounds}B), the lower bound is for all classes as good as it can be. The hypothesized upper bound, on the other hand, could certainly be improved as there are several instances for $k=2$ and for $k=3$ where it is not tight.
\end{remark}}

\begin{table}
    \centering
\includegraphics[width=0.99\textwidth]{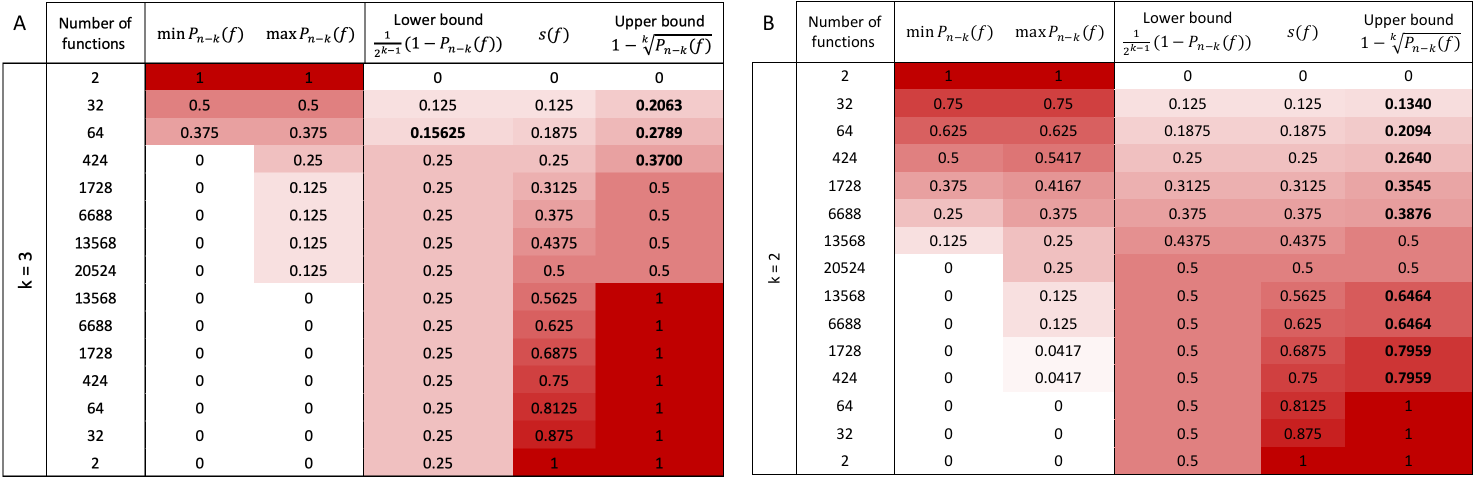} 

\caption{\new{Classification of all $2^{2^4} = 65,536$ Boolean functions in $n=4$ variables based on their average sensitivity, $s(f)$. For each class, the number of functions, minimal and maximal value of $P_{n-k}(f)$, as well as the best lower and upper bound provided by Theorem~\ref{thm_sensitivity_bounds} and Conjecture~\ref{conj} are shown. Bold font highlights bounds that are not optimal. In (A) $k=3$, i.e., $n-k=1$, in (B) $k=2$, i.e., $n-k=2$.}}
    \label{table_tight_bounds}
\end{table}

\section{Discussion}\label{sec:discussion}
Many properties of Boolean functions have been thoroughly studied over the course of the last decades. Most early studies and complexity measures of Boolean functions were motivated by questions arising from theoretical computer science. For example, Nisan used the sensitivity, the block sensitivity and the certificate complexity of a Boolean function to derive bounds for the worst-case time needed to compute a Boolean function using an ideal algorithm~\cite{nisan1991crew}. Just recently, Huang showed that all these complexity measures are polynomially related, thereby proving a major open problem in complexity theory~\cite{huang2019induced}.

The definition of the $k$-set canalizing proportions $P_k, k=0,1,\ldots,n-1$, which are the focus of this paper, is reminiscent of the definition of certificate complexity, with one big difference. Nisan defines certificates as sets of inputs to a Boolean function, which suffice to determine the output of the function~\cite{nisan1991crew}. A certificate is thus exactly a collectively canalizing input set (Definition~\ref{def:coll_canal}). The certificate complexity of an $n$-variable Boolean function, however, is defined as the number of inputs that need to be known in the \textit{worst case} (i.e., when considering all $2^n$ configurations) to determine the output of the function. The $k$-set canalizing proportion, on the other hand, quantifies the \textit{average} proportion of $k$-sets that collectively canalize a function. This highlights a general difference in the scope of use of Boolean functions in different areas of application. While theoretical computer science is particularly concerned with the worst-case scenario, the focus of biological studies is the average behavior of a system, which can be described by complexity measures like the average sensitivity of a Boolean function. 

The motivation for the complexity measures studied in this paper comes from the biological concept of canalization. Considering canalization as a property of a Boolean function (as in \cite{bassler2004evolution,reichhardt2007canalization}), rather than on the basis of individual variables as traditionally done~\cite{kauffman2003random,kauffman2004genetic,shmulevich2004activities,he2016stratification}, allows us to define and study the canalizing strength of any Boolean function. With this broader definition of canalization, we can thus distinguish finer differences in the canalization property. Given that the large majority of Boolean functions in several variables is simply not canalizing in the traditional sense (Definition~\ref{def2.3}), this constitutes a biologically relevant advancement. 

The $k$-set canalizing proportions $P_k, k=0,1,\ldots,n-1$ allow us to  connect the widely used concept of average sensitivity with the concept of canalization. Theorem~\ref{thm_sensitivity_bounds} shows how the bounds on the average sensitivity of a Boolean function tighten the more inputs we know. If we know only one input of a function (i.e., $P_1$), we can derive the number of canalizing variables. If, on the other hand, we know all but one input of a function, we can derive its average sensitivity, which is thus a bijection of the $(n-1)$-set canalizing proportion. Further, Theorem~\ref{thm_proportion_increasing} establishes bounds for the $k$-set canalizing proportions, which together with Theorem~\ref{thm_ncfs_with_one_layer}, allow us to define the most canalizing and least canalizing Boolean functions. The subsequent definition of the canalizing strength (Definition~\ref{can_strength}) yields a novel measure for how close to perfect canalization \textit{any} non-constant Boolean function is.

\bibliographystyle{unsrt}

\end{document}